\documentclass[sigplan, authorversion, nonacm]{acmart}
\usepackage[nolist]{acronym}
\usepackage{caption} 
\usepackage{subcaption}
\usepackage{tikz}
\usepackage{multirow}
\usetikzlibrary{shapes.geometric, arrows.meta, positioning}
\AtBeginDocument{%
  }

\begin{document}
\begin{acronym}
    \acro{xr}[XR]{Extended Reality}
    \acro{vr}[VR]{Virtual Reality}
    \acro{ar}[AR]{Augmented Reality}
    \acro{mr}[MR]{Mixed Reality}
    \acro{aoi}[AOI]{Areas of Interest}
    \acro{ai}[AI]{Artificial Intelligence}
    \acro{ve}[VE]{Virtual Environment}
    \acro{nlp}[NLP]{Natural Language Processing}
    \acro{svm}[SVM]{Support Vector Machine}
    \acro{hri}[HRI]{Human-Robot Interaction}
    \acro{hci}[HCI]{Human-Computer Interaction}
    \acro{ui}[UI]{User Interface}
    \acrodefplural{UI}[UI's]{User Interfaces}
    \acro{et}[ET]{Eye Tracking}
\end{acronym}
\title{The interplay of user preference and precision in different gaze-based interaction methods}


\author{Björn R. Severitt}
\email{bjoern.severitt@uni-tuebingen.de}
\orcid{0009-0000-1343-4164}
\affiliation{%
	\institution{University of Tübingen}
	\streetaddress{Maria-von-Linden-Str 6}
	\city{Tübingen} 
	\country{Germany} 
	\postcode{72076}
}

\author{Yannick Sauer}
\email{yannick.sauer@zeiss.com}
\orcid{0000-0002-7513-341X}
\affiliation{%
	\institution{Carl Zeiss Vision International GmbH}
	\streetaddress{Turnstrasse 27}
	\city{Aalen} 
	\country{Germany} 
	\postcode{73430}
}

\author{Alexander Neugebauer}
\email{a.neugebauer@uni-tuebingen.de}
\orcid{0000-0002-3254-2168}
\affiliation{%
	\institution{University of Tübingen}
	\streetaddress{Maria-von-Linden-Str 6}
	\city{Tübingen} 
	\country{Germany} 
	\postcode{72076}
}

\author{Rajat Agarwala}
\email{rajat.agarwala@uni-tuebingen.de}
\orcid{0000-0002-8036-1759}
\affiliation{%
	\institution{University of Tübingen}
	\streetaddress{Maria-von-Linden-Str 6}
	\city{Tübingen} 
	\country{Germany} 
	\postcode{72076}
}

\author{Nora Castner}
\email{nora.castner@zeiss.com}
\orcid{0000-0002-6771-7693}
\affiliation{%
	\institution{Carl Zeiss Vision International GmbH}
	\streetaddress{Turnstrasse 27}
	\city{Aalen} 
	\country{Germany} 
	\postcode{73430}
}

\author{Siegfried Wahl}
\email{siegfried.wahl@uni-tuebingen.de}
\orcid{0000-0003-3437-6711}
\affiliation{%
	\institution{University of Tübingen}
	\streetaddress{Maria-von-Linden-Str 6}
	\city{Tübingen} 
	\country{Germany} 
	\postcode{72076}
}
\affiliation{%
	\institution{Carl Zeiss Vision International GmbH}
	\streetaddress{Turnstrasse 27}
	\city{Aalen} 
	\country{Germany} 
	\postcode{73430}
}

\renewcommand{\shortauthors}{Severitt et al.}

\begin{abstract}
  In this study, we investigated gaze-based interaction methods within a virtual reality game with a visual search task with 52 participants. We compared four different interaction techniques: Selection by dwell time or confirmation of selection by head orientation, nodding or smooth pursuit eye movements. We evaluated both subjective and objective performance metrics, including NASA-TLX for subjective task load as well as time to find the correct targets and points achieved for objective analysis. The results showed significant differences between the interaction methods in terms of NASA TLX dimensions, time to find the right targets, and overall performance scores, suggesting differential effectiveness of gaze-based approaches in improving intuitive system communication. Interestingly, the results revealed gender-specific differences, suggesting interesting implications for the design of gaze-based interaction paradigms that are optimized for different user needs and preferences. These findings could help to develop more customized and effective gaze interaction systems that can improve accessibility and user satisfaction.
\end{abstract}


\maketitle

\section{Introduction}

Humans naturally rely on their gaze to perceive, explore and interact with their environment. Extending this ability to system interaction via gaze as an input method seems intuitive, because it reflects our natural focus on objects or regions of interest. However, developing a system that correctly interprets gaze is a major challenge. Systems must not only accurately capture gaze, but also interpret the user's intentions in the correct interaction syntax. There are already a number of systems that have effectively incorporated gaze input for support in immersive games~\cite{smith2006use}, smart wearables~\cite{mastrangelo2018low}, and even medical devices that support communication or mobility~\cite{pannasch2008eye,subramanian2019eye}. 
However, gaze can sometimes be pushed aside in favor of other communication approaches in human-machine interfaces. 

Effective communication in \acp{UI} often need to be designed for a specific user-need or goal. 
Often, this comes with customization challenges that can affect how intuitive the experience can be~\cite{dey2019best,macaranas2015intuitiveinteraction}. Traditional communication approaches that have evolved over the years to be more user intuitive are grounded in physical actions, w.r.t. button clicks or scrolling with a mouse, touch gestures like swiping or pinching. But the pitfalls of these devices are that they are exclusive to able-bodied individuals~\cite{gur2020human}. Verbal-based modalities are a viable alternative, especially with the recent boom in \ac{nlp} models 
. This interaction modality is increasingly preferred as it is very natural and can quickly extract the context in which the user wishes to accomplish their specific task: The best example being chatbots. Despite these advantages, \ac{nlp}-based systems face other challenges. Mainly, there are sometimes unpredictable outputs from the \emph{black box}-style of how the model learns~\cite{lin2024generatingconfidenceuncertaintyquantification, gou2024criticlargelanguagemodels}. For a user, this can result in a system being slightly annoying but also safety critical. Certain systems, for example controlling a robotic arm, can be susceptible to a user not appropriately describing the object they wish to have the robot grasp. They may not know to specify the tea kettle should be grabbed by the handle since they plan to pour it into a cup.

This challenge of understanding a user's implicit or unconscious understanding of a given context is precisely where gaze input can supplement \acp{UI}. Going back to the tea kettle example, a user would fixate momentarily on the handle moments before proceeding to other look ahead fixations indicating the next steps in the task~\cite{bovo2020detecting,Pelz2001}. This fixation input can be enough to communicate to a system, \emph{grasp here}. 
Though eye tracking technology for gaze-based communication has been around for nearly forty years, it is still a modality that is highly accessible and natural. It can be performed with relatively low-cost devices that offer high accuracy (within 1 degree) ~\cite{lee2020eyetrackinglowcost, rakhmatulin2020eyetrackinglowcostreview} and can provide scene analysis through efficient object detection methods~\cite{jha2020realtimeobjectdetection}. However, as with any input modality, there are challenges; mainly how to distinguish between an accidental gaze and a deliberate command --\emph{The Midas Touch Problem}. This issue has prompted numerous solutions, which we will overview in section~\ref{sec:related}, but when we try to understand the user experience behind these approaches, there is not enough information available.

Considering that a satisfactory user experience is important for a user's opinion of whether a \ac{ui} is useful, comfortable, or intuitive, we wanted to investigate aspects of different gaze-based interaction paradigms could contribute to the overall experience. 
In this paper, we present and evaluate four common gaze- and head-movement-based interaction methods and compare their performance in an immersive environment where participants perform a visual search task. Specifically, we investigated gaze-based selection by dwell time and methods in which gaze-based selection is confirmed by head direction, nodding, or a specific eye movement. By investigating these paradigms in terms of both objective and subjective factors, such as task performance, signal accuracy, and preference, we aim to identify interaction strategies that maximize the efficiency and usability of gaze-driven systems.

\section{Related work}
\label{sec:related}


Gaze-based interaction has been studied in detail and summarized in review articles~\cite{duchowski2018reviewgazeinteraction, plopski2022xrgazeinteractionsurvey}. Research demonstrates that gaze can be effectively leveraged for various input tasks, such as text entry~\cite{hansen2004gazetype, Majaranta2009, majaranta2002eyetype, ward2002fast, wobbrock2009textentry} and PIN entry~\cite{best2016pinentry, hoanca2006pinentry}. Furthermore, gaze-based systems have proven useful for navigating hierarchical interfaces~\cite{huckauf2007peye, huckauf2008peye}. These findings highlight the versatility of gaze as an interaction modality, making it a powerful tool for engaging with digital environments.

The most intuitive way to select an object with gaze alone is the dwell time introduced by Jacob~\cite{jacob1990dwell}. This can also be used as a metric for interest in an object~\cite{starker1990gazedwell}. As mentioned above, this method leads to the problem of the \textit{Midas Touch}, as there is no way to recognize which look is a real look of interest. To address this challenge, additional modalities such as gestures have been explored, as demonstrated by \v{S}pakov and Majaranta~\cite{spakov2012enhanced}. They investigated the use of various head gestures for different activities, such as item selection and navigation, and found that users’ preferences for gestures varied depending on the task. For selection, a nod was generally preferred, while head turning was favored for navigation, and tilting the head was most effective for switching functional modes.

Another approach involves using specific eye movements for confirmation, first introduced by Vidal et al.~\cite{vidal2013pursuits}, who employed smooth pursuit eye movements—slow, continuous motions that allow the eyes to track moving objects—to select an object. Esteves et al.~\cite{esteves2015orbits} later built on this technique, creating a spherical object that users could follow with their eyes to confirm their selection.

In \ac{vr}, head gaze and eye gaze are often compared for their effectiveness, yielding mixed results. Quian et al.~\cite{qian2017headvseye} found that head gaze performed better, while Blattgerste et al.~\cite{blattgerste2018eyevshead} reached the opposite conclusion. They attributed their findings to the higher accuracy of the eye-tracking data used in their study, which made it easier for participants to interact with the system.

A logical next step is to combine both eye and head gaze methods, as explored by Sidenmark et al.~\cite{sidenmark2019eyehead}. They proposed three different approaches for selecting a target using eye gaze, with confirmation achieved through head gaze. This combination enhances control and flexibility in the selection process. Wei et al.~\cite{wei2023headandgaze} took a different approach to the use of eye and head gaze. They created a probabilistic model based on the endpoints of the gaze and used this to decide whether and which object should be selected.

While these studies provide valuable insights into individual gaze-based and multimodal interaction techniques, most focus on evaluating a single method, often in controlled or stable environments. This approach helps to validate the feasibility of each method, but can limit understanding of how different techniques perform in more interactive or dynamic environments where users must adapt to variable conditions. In contrast, our study introduces a comparative framework that evaluates multiple gaze-based interaction methods in an immersive environment. In this way, we aim to take a broader perspective on the interaction efficiency, user experience and adaptability of different methods, thereby filling an important gap in gaze-based interaction research.

\section{Methods}

To compare the gaze-based interaction methods, we developed a custom \ac{vr} game using Unity~\cite{haas2014history}, integrated into the \textit{VisionaryVR} framework~\cite{hosp2024visionaryvr}. This setup allowed participants to test each method and then complete the NASA TLX questionnaire~\cite{hart1988nasatlx} directly afterward. Our game was designed in such a way that there is a variable search task that the participants must solve as quickly as possible in order to maximize their points (see subsection~\ref{subsec:game}). The procedure for each method began with an introduction, where the method and its adjustable parameters are explained. Then came the test phase, during which participants could try the method and adjust the parameters to their comfort. Once participants felt they had chosen their preferred parameters, the main phase could start. In the main phase, participants played the game for ten rounds. The data collected during this phase was later used for analysis. After the main phase, participants were shown the questionnaire scene to answer the NASA-TLX questions. This process was repeated for each method, with the gaze-only method always being the first to allow the participant to learn the game, and the others in random order.

\subsection{Interaction methods}

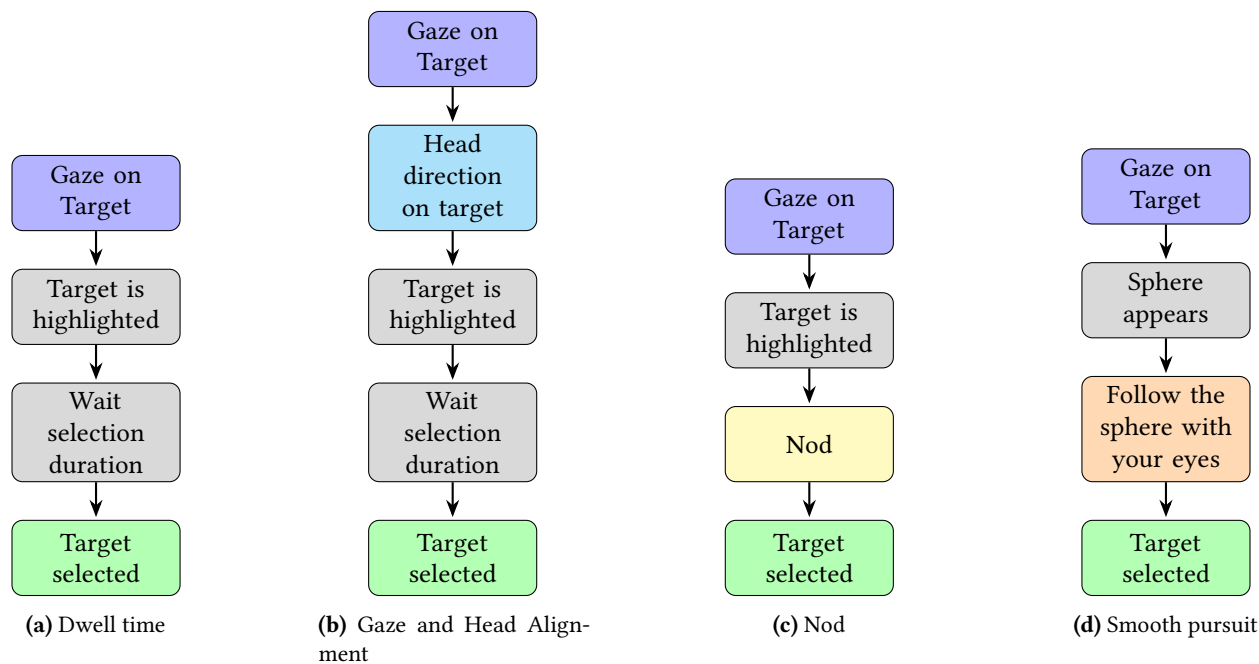
\begin{figure*}
    \centering
    \begin{subfigure}[t]{0.2\linewidth}
    \centering
        \begin{tikzpicture}[
            node distance=0.5cm and 1cm,
            box/.style={rectangle, draw, rounded corners, align=center, minimum height=1cm, text width=2cm},
            arrow/.style={-Stealth, thick, shorten >=1pt}
            ]
            
            \node[box, fill=blue!30] (Gaze) {Gaze on Target};
            \node[box, fill=gray!30, below=of Gaze] (Highlight) {Target is highlighted};
            \node[box, fill=gray!30, below=of Highlight] (wait) {Wait selection duration};
            \node[box, fill=green!30, below=of wait] (select) {Target selected};

            \draw[arrow, out=-90, in=90] (Gaze) to (Highlight);
            \draw[arrow, out=-90, in=90] (Highlight) to (wait);
            \draw[arrow, out=-90, in=90] (wait) to (select);
    
        \end{tikzpicture}
        \caption{Dwell time}
    \end{subfigure}
    \hfill
    \begin{subfigure}[t]{0.2\linewidth}
    \centering
        \begin{tikzpicture}[
            node distance=0.5cm and 1cm,
            box/.style={rectangle, draw, rounded corners, align=center, minimum height=1cm, text width=2cm},
            arrow/.style={-Stealth, thick, shorten >=1pt}
            ]
            
            \node[box, fill=blue!30] (Gaze) {Gaze on Target};
            \node[box, fill=cyan!30, below=of Gaze] (Head) {Head direction on target};
            \node[box, fill=gray!30, below=of Head] (Highlight) {Target is highlighted};
            \node[box, fill=gray!30, below=of Highlight] (wait) {Wait selection duration};
            \node[box, fill=green!30, below=of wait] (select) {Target selected};

            \draw[arrow, out=-90, in=90] (Gaze) to (Head);
            \draw[arrow, out=-90, in=90] (Head) to (Highlight);
            \draw[arrow, out=-90, in=90] (Highlight) to (wait);
            \draw[arrow, out=-90, in=90] (wait) to (select);
    
        \end{tikzpicture}
        \caption{Gaze and Head Alignment}
    \end{subfigure}
    \hfill
    \begin{subfigure}[t]{0.2\linewidth}
    \centering
        \begin{tikzpicture}[
            node distance=0.5cm and 1cm,
            box/.style={rectangle, draw, rounded corners, align=center, minimum height=1cm, text width=2cm},
            arrow/.style={-Stealth, thick, shorten >=1pt}
            ]
            
            \node[box, fill=blue!30] (Gaze) {Gaze on Target};
            \node[box, fill=gray!30, below=of Gaze] (Highlight) {Target is highlighted};
            \node[box, fill=yellow!30, below=of Highlight] (Nod) {Nod};
            \node[box, fill=green!30, below=of Nod] (select) {Target selected};

            \draw[arrow, out=-90, in=90] (Gaze) to (Highlight);
            \draw[arrow, out=-90, in=90] (Highlight) to (Nod);
            \draw[arrow, out=-90, in=90] (Nod) to (select);
    
        \end{tikzpicture}
        \caption{Nod}
    \end{subfigure}
    \hfill
    \begin{subfigure}[t]{0.2\linewidth}
    \centering
        \begin{tikzpicture}[
            node distance=0.5cm and 1cm,
            box/.style={rectangle, draw, rounded corners, align=center, minimum height=1cm, text width=2cm},
            arrow/.style={-Stealth, thick, shorten >=1pt}
            ]
            
            \node[box, fill=blue!30] (Gaze) {Gaze on Target};
            \node[box, fill=gray!30, below=of Gaze] (Sphere) {Sphere appears};
            \node[box, fill=orange!30, below=of Sphere] (Follow) {Follow the sphere with your eyes};
            \node[box, fill=green!30, below=of Follow] (select) {Target selected};

            \draw[arrow, out=-90, in=90] (Gaze) to (Sphere);
            \draw[arrow, out=-90, in=90] (Sphere) to (Follow);
            \draw[arrow, out=-90, in=90] (Follow) to (select);
    
        \end{tikzpicture}
        \caption{Smooth pursuit}
    \end{subfigure}
    \caption{Summary of the gaze-based selection methods, each method is distinguished by the purple colored boxed. All methods start by looking at a target. The gray boxes indicate the interface actions in response to the user input for each of the methods. Then, a confirmation is necessary (blue, yellow and orange), aside from the method in (a). }
    \label{fig:SummarySelectionMethods}
\end{figure*}

Each selection method uses gaze as the core interaction technique, though we investigated differing modalities to confirm a user's selection. Below, we describe each method in detail and how participants could customize them to their preferences (see the additional Fig.~\ref{fig:SummarySelectionMethods} depicting each method).

\textit{Gaze Dwell.}
This method is the most common and is based exclusively on the dwell time introduced by Jacob~\cite{jacob1990dwell}. Here, a target is selected by fixating on it for a sufficient duration. As the player gazes at an object, the target is highlighted with an outline. We implement gaze dwell as follows. The time spent looking at the target is visually represented by a change in the outline color, transitioning from green to red as the selection is locked in. For this method, the only parameter the participant must set is the \textit{selection duration}, which specifies the time in seconds the user must gaze at the target for it to be selected. We set a minimum duration to 0.05 and a maximum duration to 1 and the step increments were 0.05.

\textit{Gaze and Head.}
This method is inspired by the method from  Sidenmark et al.~\cite{sidenmark2019eyehead}. Here, the player must not only look at the target, but also confirm the selection by aligning their head with the target. The direction of the head was indicated by a green dot. The target is highlighted with a magenta outline when gaze is directed towards it, and with a green outline when head and gaze are directed to it. The green border changes to red to visualize the dwell time. For this method, the participant must set two parameters: (1) \textit{selection duration} is similar as for the \textit{dwell time} method, but now  it represents the duration the gaze and the head direction have to be aligned. (2) \textit{Head Orientation Precision} specifies the allowed offset in degree of the user's head orientation relative to the target for the selection confirmation. We set a minimum degree to 2 and a maximum degree to 20 and the step increments were 2. Thus, a higher degree indicates less precision the system needs to confirm selection.

\textit{Nod.}
This method is based on the results of \v{S}pakov and Majaranta~\cite{spakov2012enhanced}. The nodding method involves selecting the target by first gazing at it and then performing a nod gesture. The nod is recognized as a confirmation signal to complete the selection. For this method, the participant must set two parameters when the gesture is recognized as a nod: (1) \textit{nod strength}, which specifies the amount of head movement in degree required for the system to register the nod gesture. (2) \textit{Nod direction precision}, which specifies the allowed offset in degree of the final head position required to confirm the nod. For (1), we set a minimum degree to 5 and a maximum degree to 30 and the step increments were 1. For (2), we set a minimum degree to 1 and a maximum degree to 20 and the step increments were 1. 

\textit{Smooth Pursuit.}
This method follows the idea of Esteves et al.~\cite{esteves2015orbits}. For our implementation, an orange-colored sphere appears after looking at the target and starts to move along a pre-defined path. The player must follow the sphere's movement with their gaze. If the player tracks the sphere  and the smooth pursuit's trajectory and velocity align to the moving sphere, the target will be selected. This method has two different parameters to set: (1) \textit{Tracking precision}, which represents the required correlation between the user's viewing direction and the movement of the sphere. We have set a minimum of 0.05 and a maximum of 1, with an increment of 0.05.
(2) \textit{Movement pattern}, which is the path that the sphere follows. Here the participant can choose between three options: \textit{Circle}, \textit{Bounding} and \textit{Random Walk}.

\subsection{Game} \label{subsec:game}

We evaluated each interaction method in the \ac{vr} game environment that we developed. The goal was to get a high score. To accomplish this, they have to tweak the parameters for each interaction method to give them the best performance. Thus, they would have to figure out factors such as comfort, speed, and accuracy that would help them achieve the most points in the allotted time.

\textit{Game environment.}
The game takes place in a square-shaped room that contains several interactive elements distributed across its walls. One wall displays the high score list, showing the top 9 scores. Another wall shows the current score, which resets at the beginning of each new round. A third wall presents the remaining time for the current round. The final wall contains a set of sliders that allow players to adjust the parameters of the interaction method in use. 

\textit{Gameplay.}
The core mechanics of the game revolve around flying robots. The robots are randomly scattered throughout the room and could occlude each other.  As soon as the robots have reached their position in the room for the round, the target or distractor appear in a sphere that makes up their body: The target is the letter "C" and the distractors are the letter "O".Players must identify and destroy the correct target using the specific gaze selection method for the current experimental block. Properly selecting the target from the distractors gives no penalty, whereas improperly selecting the distractor results in points subtracted from the score. When the correct target is destroyed, the robots fly around to new positions and a new target has to be found. This continues for 30 seconds, then the round is over. The four experiment blocks are rounds related to each gaze selection method.
See the supplementary video for a playback of the gameplay with each of the selection methods.

\textit{Scoring System.}
Points are awarded based on the player's performance according to the following rules:
Positive points range from 5 to 20 for successfully destroying the correct target. If the player takes more than 10 seconds to destroy the target, they will only receive the minimum of 5 points. Otherwise, the points are determined through linear interpolation, with faster responses yielding higher scores.
Minus points are awarded if an incorrect target is destroyed or a robot is shot at, resulting in a deduction of 21 points. This choice for scoring adds an additional time pressure to the participant. 

\subsection{\ac{vr} setup}

The game was conducted in \ac{vr}. Participants interacted with the system using the HTC Vive Pro Eye (HTC Corporation, Taoyuan, Taiwan), which includes a built-in Tobii eye tracker (Core SW 2.16.4.67) with an estimated accuracy of $0.5^\circ - 1.1^\circ$ and a sampling frequency of 120 Hz. Eye tracking data was calibrated and accessed via the Vive SRanipal SDK (HTC Corporation, Taoyuan, Taiwan)~\cite{sipatchin2021vrheadset}. In addition, the participants were provided with an HTC Vive Pro Controller 2.0 to adjust the slider for the parameters for each interaction method and to answer the NASA-TLX questions in the questionnaire scene.

To extract gaze data from the \ac{vr} headset, we employed the ZERO-Interface~\cite{zero2023hosp}, which is integrated into VisionaryVR. This interface provides separate three-dimensional gaze vectors for each eye, along with a combined gaze vector. The data are accessible in real time, allowing them to be used for both gameplay and executing interaction methods. Additionally, all gaze data were recorded for further analysis.

\subsection{Participants}

\begin{table*}[]
    \centering
\begin{tabular}{lrlrrrr}
\toprule
{} &  Count & Mean age &  Minimal age &  Maximal age &  VR Experience &  ET Experience \\
\midrule
Woman       &     32 &    26.31 &           18 &           63 &             15 &             11 \\
Man         &     18 &    27.56 &           18 &           67 &             10 &              2 \\
Non-binary    &      1 &     24.0 &           24 &           24 &              0 &              1 \\
Prefer Not to Say  &      1 &     23.0 &           23 &           23 &              0 &              0 \\
\hline
General      &     52 &    26.63 &           18 &           67 &             25 &             14 \\
\bottomrule
\end{tabular}
    \caption{Demographic and experience data of study participants, including total number, average age, age range (minimum and maximum age) and experience with \ac{vr} and \ac{et} technologies. The data is presented in total and broken down by gender identity to provide insights into the diversity of participants and corresponding technological familiarity.}
    \label{tab:PatMetaData}
\end{table*}

52 people took part in this study. These people came from the University of Tübingen and were mainly students. 32 of them self-identified as women, 18 identified as men, one identified as non-binary, and one preferred not to provide gender information. Table~\ref{tab:PatMetaData} also shows the age range and how many had experience with \ac{vr} and \ac{et}. In this case, experience meant that the participants had at some point used a device with eye tracking capabilities. All participants were given a brief introduction to the gameplay, similar to subsection~\ref{subsec:game}, and learned how to use the controller to adjust the settings and start a round.

The studies involving human participants were reviewed and approved by Faculty of Medicine at the University of Tübingen with a corresponding ethical approval identification code 986/2020BO2. The participants provided their written informed consent to participate in this study.

\subsection{Measurements}

Several measurements were carried out to evaluate and compare the methods. In addition to the NASA-TLX questionnaire, participants were asked which method they would prefer if they had to choose one. Objective measures, such as score, were also recorded for quantitative comparisons. Participants could change settings as often as they wished, but each gaze selection method required at least ten consecutive rounds at their fixed settings. Objective measurements were only taken in these ten rounds to determine the most comfortable settings for each participant.

\textit{NASA-TLX}.
That questionnaire was created to measure the task load of a participant. It is widely used and has six different dimensions: \textit{Mental Demand} evaluates how much mental and perceptual effort a task requires. The participants were asked, how much of thinking and deciding they had to do. \textit{Physical Demand} assesses the physical effort needed to complete the task, which included movement. \textit{Temporal Demand} considers the time pressure to which the participant was exposed and determines whether they perceived the pace of work as hurried or leisurely. \textit{Performance} evaluates how the participants felt the successfulness of the completion of the task. \textit{Effort} measures the amount of physical and mental energy that the participants believe they had to expend to solve the task. \textit{Frustration} analyses emotional stress and takes into account factors such as anger, uncertainty, and frustration during the task. For each dimension, participants rate on a scale from low to high, except for performance, where it ranges from perfect to failure.

\textit{Objective measurements}. 
In addition to the subjective ratings, several objective characteristics were measured directly from the data: \textit{Time on Task} refers to the time it took participants to select the correct target. \textit{Points} represent the scores that participants achieved in each round. \textit{Fails} indicates the number of incorrect selections, i.e. selecting a sphere with the character "O" instead of the "C". Therefore, we have the number of points and errors per round of a participant, and the time for the task depends on the number of correct selections.
\section{Results}

We first evaluate the subjective measurements, namely participant preferences and NASA-TLX responses regarding interaction method. We then follow with the objective measurements like task duration and performance related to interaction method. As we are interested in which settings they preferred for each interaction method, we then compare the distributions of each setting. 
Since the data are not normally distributed and do not have the same variance, we chose non-parametric tests such as Kruskal~\cite{kruskal2010mckight} to check if there are differences between several groups and Mann-Whitney U~\cite{mannwhitneyu2010mcknight} tests to compare two specific groups.

\subsection{Subjective measurements}

\begin{table*}[]
    \centering
\begin{tabular}{lrrrr}
\toprule
{} &  Gaze Dwell &  Head and gaze &  Nod &  Smooth pursuit \\
\midrule
Woman       &    10 &              6 &   13 &               3 \\
Man         &     2 &              8 &    8 &               0 \\
NonBinary    &     1 &              0 &    0 &               0 \\
Prefer Not to Say &     0 &              1 &    0 &               0 \\
\hline
General      &    13 &             15 &   21 &               3 \\
\bottomrule
\end{tabular}
    \caption{The preferences of the participants for different methods, indicating the number of people who selected each method as their preferred method. Preferences are categorized by gender identity for the four methods.}
    \label{tab:PrefMethod}
\end{table*}

Overall, the \textit{Nod} method proved to be the most favored selection method by the participants, followed by the \textit{head and gaze} and \textit{Gaze Dwell} methods, which showed a similar level of popularity. Table~\ref{tab:PrefMethod} details the preferences for each of the interaction methods. We observed unexpected differences in preferences across gender categories, especially for the \textit{Gaze Dwell} method, which is predominantly favored by women participants: Ten out of a total of 13 who preferred this method. In contrast, men participants showed equal preference between the methods \textit{Head and Gaze} and \textit{Nod}, with 8 participants preferring each of these options. The \textit{Smooth Pursuit} method showed a minimal preference in all groups, with only three participants - all women - choosing this method.
Overall, these results emphasize the broad appeal of the \textit{Nod} method while showing clear gender preferences within other methods.

\begin{figure}
    \centering
    \includegraphics[width=\linewidth]{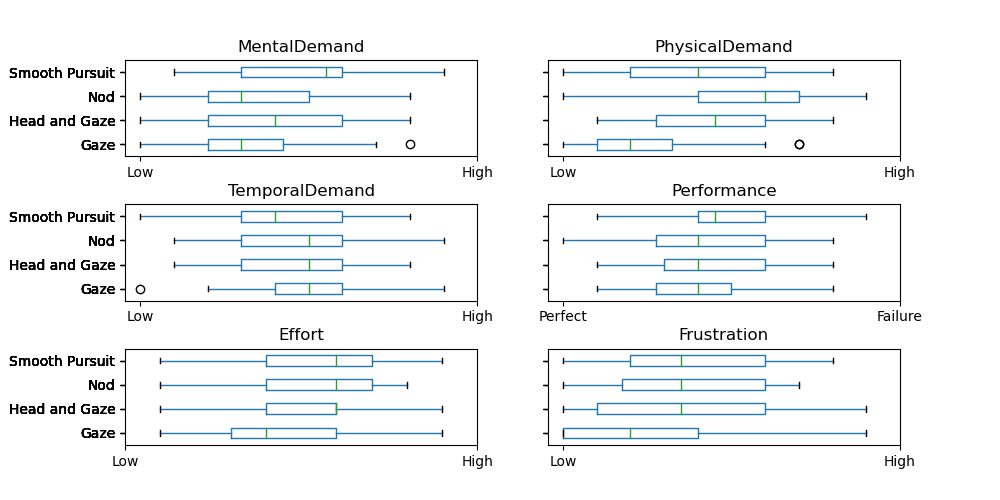}
    \caption{Boxplots summarizing NASA-TLX scores across six dimensions for each of the interaction methods, showing the distribution of responses. \emph{Gaze} on the y-axis is the shortened version of Gaze Dwell}
    \label{fig:AnswersNASATLX}
\end{figure}

For some of the NASA-TLX dimensions, there were slight overall differences between the methods (see Fig.~\ref{fig:AnswersNASATLX}). In particular, there are statistically significant differences in the dimensions of Mental Demand, Physical Demand, Effort and Frustration, with \textit{Smooth Pursuit} and the other gaze-based interaction methods showing significant differences across multiple measures. We report these in detail below.

\textit{Mental Demand}. The methods show significant differences between the groups in these dimensions ($H=10.694$, $p=0.014$). The post hoc Mann-Whitney U test showed that the \textit{Smooth Pursuit} method is significantly different from the other methods ($p<0.05$) and with the information from Fig.~\ref{fig:AnswersNASATLX} we see that it is higher. No significant differences were found for the other methods ($p>0.05$).

\textit{Physical Demand}. The physical demand dimension has a significant difference between the methods ($H=29.720$, $p<0.001$). The \textit{Gaze Dwell} method is particularly low, while the others lead to a higher physical demand, with \textit{Nod} reaching the highest value. Only between \textit{Smooth Pursuit} and \textit{Head and Gaze} no significant difference was found ($p=0.170$).

\textit{Temporal Demand}. No statistically significant differences were found between the methods in this dimension ($H=6.502$, $p=0.09$). However, it is noteworthy that the median temporal demand in the \textit{Smooth Pursuit} method is lower compared to the other methods, indicating a possible tendency for this approach to cause less time pressure. 

\textit{Performance}. No significant differences were found in this dimension either ($H=7.257$, $p=0.064$). It is noteworthy that the \textit{Smooth Pursuit} method tends to perform below average compared to the other methods, which indicates a possible tendency towards lower effectiveness, especially compared to the gaze and head nodding methods.

\textit{Effort}. There are significant differences between the methods based on the Kruskal test ($H=11.308$, $p=0.010$). The post hoc test showed that only the gaze differs significantly from the other methods ($p<0.05$). The effort is therefore significantly lower than with the other methods. The other methods showed no significant differences ($p>0.05$).

\textit{Frustration}. This is similar to the effort dimension. There are significant differences between the methods ($H=10.917$, $p=0.012$), but these are between Gaze Dwell and the others. The perceived frustration is lower with the Gaze dwell method than with the other methods.

\subsection{Objective measurements}

\begin{figure}
    \centering
    \includegraphics[width=\linewidth]{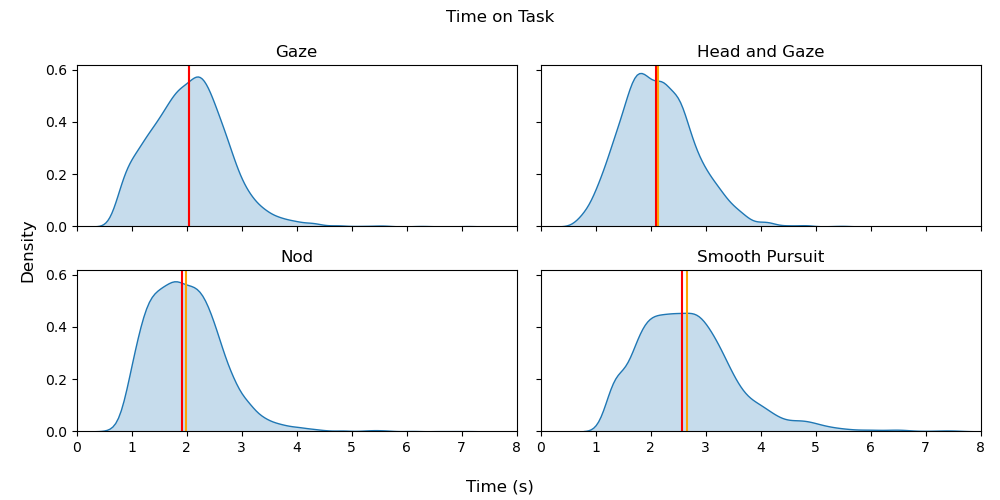}
    \caption{Distribution of the time required by the participants to find and select the target for each interaction method. The orange line indicates the mean and the red line the median.}
    \label{fig:TimeOnTask}
\end{figure}

Fig.~\ref{fig:TimeOnTask} shows the distribution of the time the participants take to destroy the target. There are significant differences between the interaction methods 47 ($H=1803.864$, $p<0.001$). A post hoc Mann-Whitney-U test with adjusted p-values using the Bonferroni method shows that these differences are between all groups ($p<0.001$).

\begin{table*}[]
    \centering
\begin{tabular}{lrrrrr}
\begin{tabular}{lrrrrr}
\toprule
{} &  Mean Time &  Std Time &  Fails &  Mean Points &  Std Points \\
\midrule
Gaze  Dwell        &      2.049 &     0.716 &        171 &      144.626 &      32.131 \\
Head and Gaze   &      2.137 &     0.678 &         73 &      147.737 &      25.573 \\
Nod           &      1.985 &     0.691 &        114 &      159.578 &      29.265 \\
Smooth Pursuit &      2.665 &     0.905 &         21 &      122.810 &      23.150 \\
\bottomrule
\end{tabular}
\end{tabular}
    \caption{Summarized performance metrics for the methods. The mean time is the average time it takes participants to find the correct target. The number of incorrect targets is the number of incorrectly selected targets over all rounds of the methods and the points are the average value over all rounds.}
    \label{tab:PerformanceSummary}
\end{table*}

\begin{figure}
    \centering
    \includegraphics[width=\linewidth]{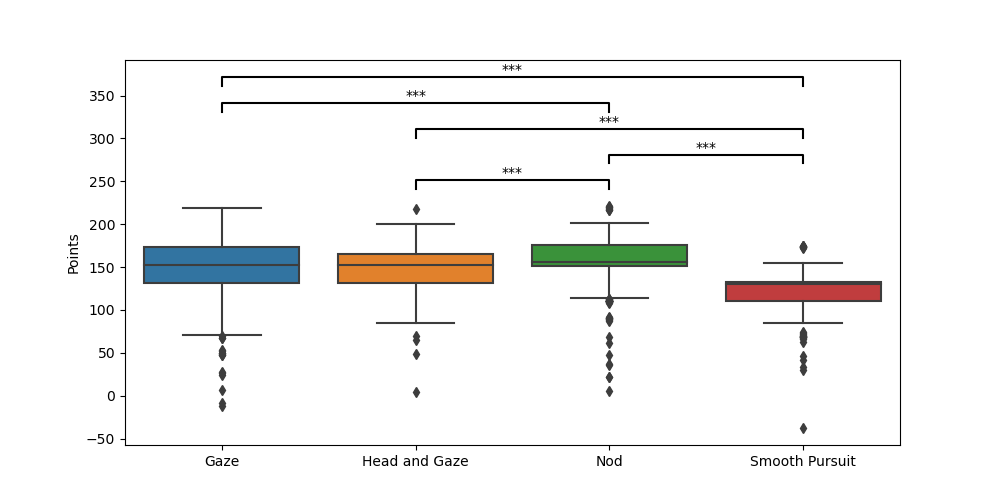}
    \caption{Boxplot summary of the points obtained with the different interaction methods. Significant differences between the methods are labelled with ‘***’ to indicate $p < 0.001$.}
    \label{fig:PointsBoxplot}
\end{figure}

Table~\ref{tab:PerformanceSummary} shows summarized performance metrics for each method, including the average score, the time to find the correct target, and the number of incorrect target choices across all rounds. Significant differences in scores between methods are observed ($K=550.302$, $p<0.001$). 
The significant differences between the methods are illustrated in Fig.~\ref{fig:PointsBoxplot}. \textit{Gaze Dwell} and \textit{Head and Gaze} score similar points, while \textit{Nod} scores were the highest and \textit{Smooth Pursuit} scores were the lowest.

\subsection{Gender differences}

Since preference for the interaction method suggests that gender has a potential influence, we decided to further examine whether gender differences were also apparent in NASA-TLX questionnaire results. Overall, we observed no significant differences between genders across the NASA-TLX dimensions ($p>0.05$). 
As, we only had two participants who identified as non-binary and preferred not to disclose this information, we do not wish to over-generalize their feedback.  

\begin{figure}
    \centering
    \includegraphics[width=\linewidth]{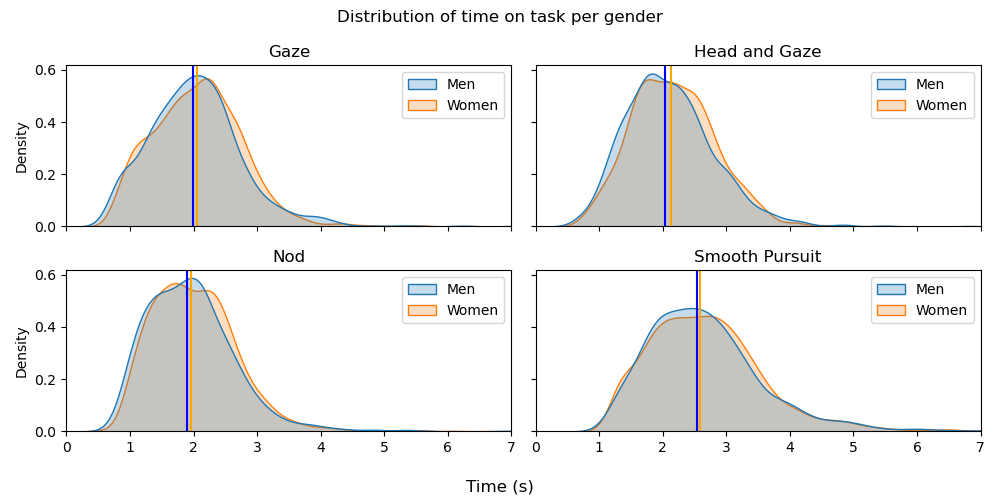}
    \caption{Distribution of time spent on a task, broken down by gender. The vertical lines indicate the median value.}
    \label{fig:timeOnTaksGender}
\end{figure}

Fig.~\ref{fig:timeOnTaksGender} illustrates the distribution of the time required to select the correct targets, separated by gender. The total time to find the correct target shows only small differences between the genders and the different methods. However, significant differences are observed for all methods, except for \textit{Smooth Pursuit}, where times did not differ significantly between genders (Mann-Whitney U-test, $p<0.05$ for other methods). This suggests that although selection times are generally comparable, subtle but statistically significant gender differences can be observed for certain methods.

\begin{table}[]
    \centering
    \begin{tabular}{llrrr}
\toprule
Method          & Gender & Points & Time (s) & Error Rate \\
\midrule
\multirow{2}{*}{Gaze Dwell}    & Men   & 144.23 & 2.03 & 3.11 \\
                               & Women & 144.17 & 2.06 & 3.53 \\
\multirow{2}{*}{Head and Gaze} & Men   & 148.47 & 2.10 & 1.61 \\
                               & Women & 146.20 & 2.17 & 1.34 \\
\multirow{2}{*}{Nod}           & Men   & 159.38 & 1.98 & 3.11 \\
                               & Women & 158.72 & 2.03 & 1.69 \\
\multirow{2}{*}{Smooth Pursuit}& Men   & 121.70 & 2.68 & 0.78 \\
                               & Women & 122.86 & 2.67 & 0.19 \\
\bottomrule
\end{tabular}
    \caption{Performance metrics by gender for all four methods, showing mean points scored, mean time to select the correct target and total number of failures per participant across all rounds.}
    \label{tab:GenderDiff}
\end{table}

Table~\ref{tab:GenderDiff} presents performance metrics by gender. While no significant differences are found in points scored (all $p>0.05$), significant differences are apparent in the time taken to select the correct target. Men generally make more incorrect selections than women, particularly with the \textit{Nod} method ($U=32927.0$, $p=0.015$), and also show slightly higher error rates in \textit{Smooth Pursuit} ($U=32752.5$, $p=0.004$). For \textit{head and gaze}, men also show a higher error rate, but this does not differ significantly from women error rates ($U=34486.0$, $p=0.164$). In contrast, only with the \textit{Gaze Dwell} method do women make more errors than men, but this was not significant ($U=29251.5$, $p=0.493$).

\begin{figure}
    \centering
    \includegraphics[width=\linewidth]{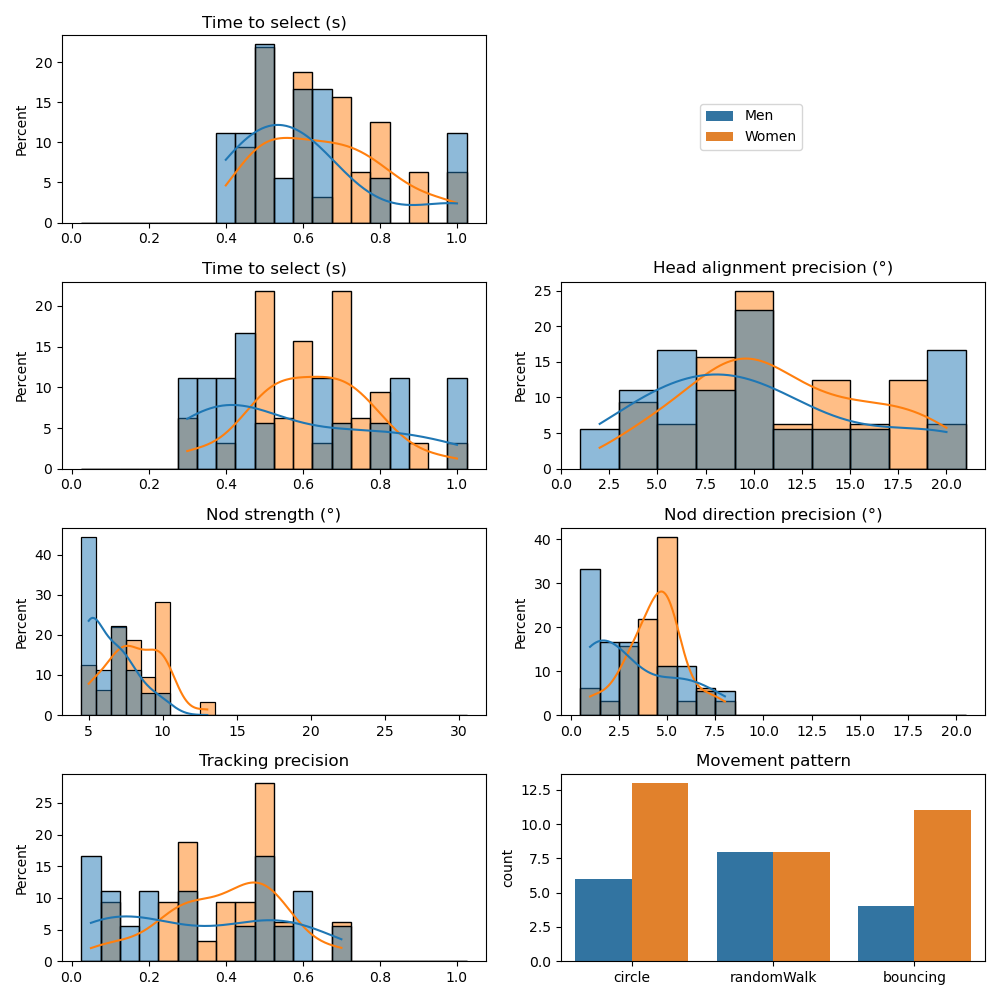}
    \caption{Histogram of selected parameters per gender. The first line is for the \textit{Gaze Dwell} method, the second line for the \textit{Head and Gaze} method, the third for the \textit{Nod} method and the last for \textit{Smooth Pursuit}.}
    \label{fig:ChoosenParameter}
\end{figure}

Fig.~\ref{fig:ChoosenParameter} shows the distribution of the parameters used, broken down by gender. Statistically significant differences between the genders are only found for the Nod parameters, especially for \textit{Nod Strength} ($U=143.0$, $p=0.003$) and \textit{Nod Direction Precision} ($U=191.5$, $p=0.048$). For the \textit{Time to Select} settings, no significant differences are observed between the \textit{Gaze Dwell} and \textit{Head and Gaze} methods, for both men ($U=182.5$, $p=0.525$) and women participants ($U=559.0$, $p=0.528$). This emphasizes that, apart from the nod parameters, the selection settings are generally consistent across genders.

\subsection{Influence of preferrences}

\begin{figure}
    \centering
    \includegraphics[width=\linewidth]{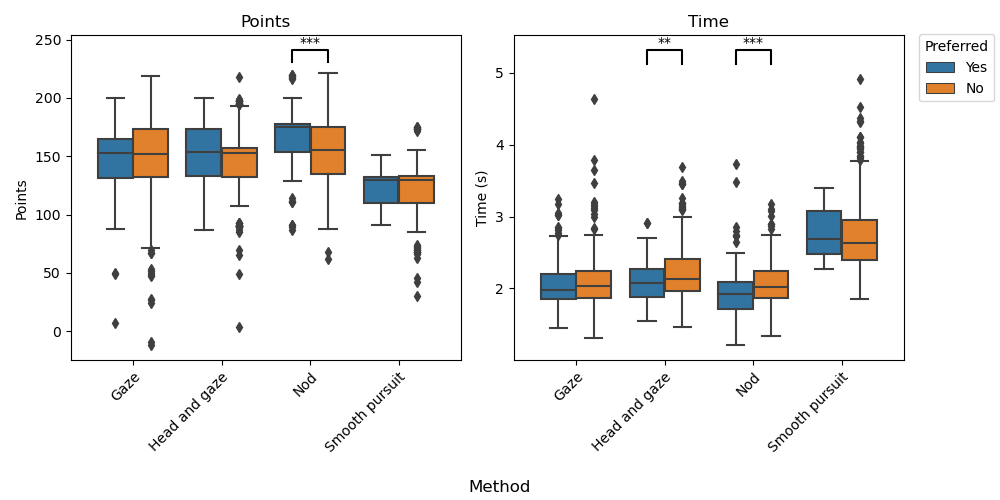}
    \caption{Boxplots showing the distribution of points scored and completion times for participants who preferred each method compared to those who did not. Significant differences are labeled with '**' to indicate $p<0.01$ and '***' to indicate $p<0.001$.}
    \label{fig:PrefYesNo}
\end{figure}

Fig.~\ref{fig:PrefYesNo} shows that objective measures, such as the number of points achieved and the processing time, do not differ significantly between participants who favored a method and those who did not. However, there is a significant difference in the scores for the \textit{Nod} method, while the processing time for the \textit{Head and Gaze} and \textit{Nod} methods is significantly different.

Further analysis of the NASA-TLX responses revealed no significant differences between participants who favored a method and those who did not. This suggests that subjective perceptions of workload, such as mental and physical demands, effort and frustration, are relatively consistent regardless of the preferred method. 

\section{Discussion}
For this study, we developed a \ac{vr} game environment to evaluate and compare four common gaze-based interaction methods, focusing on subjective and objective measures of user experience and performance. Using NASA-TLX questionnaire, subjective workload was assessed across six dimensions, while objective measures such as time on task, score, and error rate provided quantitative insights. Participants adjusted interaction settings for optimal comfort, and results were analyzed for possible gender differences in preferences and performance. This comprehensive approach aimed to identify the strengths and weaknesses of each method and provided valuable insights for designing more intuitive and user-friendly gaze-based interfaces.

The NASA-TLX results show significant differences in several dimensions between the interaction methods. The smooth pursuit for target selection in particular showed a higher mental demand than other methods, indicating a potentially more cognitively intensive interaction. The physical demand varied significantly, with the gaze method requiring the least physical effort. 
Although the overall differences in time requirements were not statistically significant. 
Performance ratings showed only marginal differences, with smooth pursuit tending to lead to lower perceived success, especially compared to gaze dwell and nodding movements. Effort and frustration scores were significantly lower for the gaze dwell method compared to the other methods, which can be attributed to the ease of use and lower frustration scores. More importantly, these results were consistent between genders and there were no significant gender differences in perceived workload. These results emphasize the different cognitive and physical profiles of each method, with the gaze dwell method proving to be the most user-friendly in terms of physical and emotional demands, while the smooth pursuit method had higher demands on mental and time resources.

The performance metrics show significant differences between the interaction methods in terms of time, accuracy and score. The nod method achieved the highest average score, while the smooth pursuit method had the lowest average score but resulted in the fewest incorrect selections, indicating a trade-off between speed and accuracy. The gaze dwell time and head and gaze methods achieved comparable results, although the gaze dwell time method produced the highest number of incorrect selections. The average time to complete the task varied considerably, with the smooth pursuit method taking the longest time overall. These results highlight clear performance conflicts where the nod method maximizes scoring potential, but the smooth pursuit is more error resistant.

Regarding preference, the nod interaction method was the most preferred in general, with men preferring mainly the head and gaze or nod methods and women preferring the gaze dwell or nod methods. However, the analyses of the NASA-TLX questionnaire show no significant differences in the perceived task load between the genders. There are significantly differences in the objective measurements, but these are so small that they are not recognized in reality. This suggests that the observed preferences for the gaze dwell method are not due to differences in task demands or workload. Instead, they could stem from other factors, such as individual familiarity with gaze-based interactions, comfort level or specific engagement with the game mechanics, suggesting a more nuanced understanding of the influence of gender on interaction preferences.

The gender-specific differences in selection time and scoring are minimal, with both groups showing similar performance across all methods. However, we found that men tend to make more selection errors than women, particularly  when using the nod and the smooth pursuit methods. Significant gender-specific differences in the selected parameter preferences only occur for the settings \textit{Nod Strength} and \textit{Nod Direction Precision}. The selection settings, such as \textit{Time to Select} for all gaze-based methods, is consistent for all genders. These results suggest that, while performance is broadly comparable, there are subtle differences in error rates.

Interestingly enough, though target selection through dwell time has been one of the most common gaze-based interaction methods \cite{namnakani2023comparedwell, Chen2017ImprovingGS} to combat the \emph{Midas Touch} problem, we found other methods were only slightly more preferred. We attribute this to the structure of the game, which requires frequent and rapid target selection, the participants' gaze is constantly in an intentional target selection mode, with little chance for unintentional eye movements, reducing the effect of the \textit{Midas Touch} problem \cite{hyrskykari2012gazeinteraction}. Although the problem is less apparent compared to a more naturalistic task, there were still error rates, especially for the gaze dwell method. This suggests that while gaze dwell provides an intuitive and fast selection process, the associated unintended activations can lead to increased inaccuracies, especially in fast or continuous selection tasks. Potentially more naturalistic tasks may have users preferring other methods even more.

\section{Potential implications for society}
Our research aim was initially focused on how user preference for different gaze-based interaction paradigms related to personal opinions of demand and comfort, and how this could affect task performance factors. We did find that subjective comfort can sometimes come with a tradeoff of accuracy. Even more so, we did not expect gender to become such a relevant factor in our analyses. Therefore, we feel this research does have potential implications for ethical aspects, namely fairness. We suggest that designing user interfaces that employ gaze should not only understand the importance of customizability and personal preference, but a user's choice could be affected by other factors such as gender, potentially diversity, or other sociodemographic. By no means are we implying digital tools for specific genders should be different -- for instance, less precise and more pink for women -- rather encouraging a broader perspective when developing these tools for varied use cases.

\section{Limitations}

While our study provides valuable insights into gaze-based interaction methods, it is important to note that these results are from a gaming environment. In such an environment, convenience and ease of interaction often take precedence over strict efficiency, as users are generally more tolerant of occasional incorrect choices if it improves their overall experience. This prioritization can differ significantly in real life or professional applications, where accuracy and efficiency are often paramount. Therefore, the trade-off between effort and efficiency observed in our study may not directly translate to contexts other than games. In practical applications where the stakes are high, users may favor precise control and minimal errors over convenience, shifting the balance of these factors. Future research should test these methods in scenarios with practical, real-world requirements to better assess their broader applicability.
\section{Conclusion}

This study investigated different gaze-based and combined gaze-head interaction methods in an interactive game environment and examined how these methods affect performance and error rates. Our results show significant differences between the methods in terms of perceived workload and accuracy, suggesting that some approaches provide intuitive and comfortable interaction, while others have higher precision. Gender differences were also found, with preferences and performance differing for certain methods, suggesting that interaction systems should take individual differences into account by offering customizable, personalisable options.

While gaze-only methods remain popular, combined gaze-head approaches show the potential for more accurate target selection with fewer errors. However, in a gaming environment, precision may not be as important to the participant, as errors have less severe consequences than in the real world, which may mean that these results are different in real-world applications.
This emphasis the importance of tailoring interaction methods to specific use cases, taking into account both preferences and performance requirements. Future research should further investigate these methods in different environments to test their practicality and inclusivity in real-world scenarios.

\begin{acks}
This research is supported by European Union's Horizon 2020 research and innovation program under grant agreement No.951910 and the German Research Foundation (DFG): SFB 1233, Robust Vision: Inference Principles and Neural Mechanisms, TP TRA, project No. 276693517.
\end{acks}

\bibliographystyle{ACM-Reference-Format}
\bibliography{Content/references}

\end{document}